%% file: oumuamua_gdr2.tex
\documentclass[onecolumn]{aastex62}
\title{1I-Home}
\usepackage{amsmath}
\usepackage{amssymb}
\usepackage{mathtools}
\usepackage{multirow}
\usepackage{rotating}

\def\myeol{\\}

\newcommand{\okina}{`}

\providecommand{\oum}{{{\okina}Oumuamua}}
\newcommand{\Hawaii}{Hawai{\okina}i}

\providecommand{\vinf}{\ensuremath{v_\infty}}
\providecommand{\candsaa}{2\,k=2}
\providecommand{\candsbb}{3\,k=2}
\providecommand{\candscc}{3\,k=1}
\providecommand{\candsdd}{7d}
\providecommand{\candsee}{7e}
\providecommand{\candsff}{7c}

\providecommand{\gaia}{Gaia}
\providecommand{\gdr}[1]{Gaia\,DR{#1}}
\providecommand{\gmag}{\ensuremath{G}}
\providecommand{\mg}{M$_\gmag$}
\providecommand{\bprp}{BP-RP}
\providecommand{\teff}{\ensuremath{T_{\rm eff}}}

\providecommand{\tenc}{\ensuremath{t_{\rm enc}}}
\providecommand{\denc}{\ensuremath{d_{\rm enc}}}
\providecommand{\venc}{\ensuremath{v_{\rm enc}}}
\providecommand{\tenclma}{\ensuremath{t_{\rm enc}^{\rm lma}}}
\providecommand{\denclma}{\ensuremath{d_{\rm enc}^{\rm lma}}}
\providecommand{\venclma}{\ensuremath{v_{\rm enc}^{\rm lma}}}
\providecommand{\tencmed}{\ensuremath{t_{\rm enc}^{\rm med}}}
\providecommand{\dencmed}{\ensuremath{d_{\rm enc}^{\rm med}}}
\providecommand{\vencmed}{\ensuremath{v_{\rm enc}^{\rm med}}}

\providecommand{\ra}{\ensuremath{\alpha}}
\providecommand{\dec}{\ensuremath{\delta}}

\providecommand{\parallax}{\ensuremath{\varpi}}
\providecommand{\parzp}{\ensuremath{\varpi_{\rm zp}}}

\providecommand{\propm}{\ensuremath{\mu}}
\providecommand{\vx}{\ensuremath{v_x}}
\providecommand{\vy}{\ensuremath{v_y}}
\providecommand{\vz}{\ensuremath{v_z}}
\providecommand{\sigmavx}{\ensuremath{\sigma(\vx)}}
\providecommand{\sigmavy}{\ensuremath{\sigma(\vy)}}
\providecommand{\sigmavz}{\ensuremath{\sigma(\vz)}}
\providecommand{\corvxvy}{\ensuremath{\rho(\vx, \vy)}}

\renewcommand{\vr}{\ensuremath{v_r}} 

\providecommand{\kms}{\ensuremath{\textrm{km\,s}^{-1}}}
\providecommand{\maspyr}{\ensuremath{\textrm{mas\,yr}^{-1}}}
\providecommand{\msun}{\ensuremath{M_\odot}}
\providecommand{\degree}{\ensuremath{^\circ}}

\usepackage{xcolor}
\usepackage{url}
\usepackage{longtable}
\usepackage{hyperref} 
\definecolor{VeryDarkBlue}{RGB}{0,0,80}
\definecolor{VeryDarkRed}{RGB}{90,0,00}
\hypersetup{ colorlinks = true, 
 urlcolor = VeryDarkBlue, 
 linkcolor = VeryDarkBlue,  
 citecolor = VeryDarkRed 
}

\received{13 August 2018}
\revised{17 September 2018}
\accepted{23 September 2018 to AJ}

\shorttitle{Plausible home stars of \oum}
\shortauthors{Bailer-Jones {\it et al.}}

\begin{document}

\title{Plausible home stars of the interstellar object \oum\ found in \gdr{2}}

\author{Coryn A.L.\ Bailer-Jones}
\affil{Max Planck Institute for Astronomy, K\"onigstuhl 17, 69117 Heidelberg, Germany}

\author[0000-0003-0774-884X]{Davide Farnocchia}
\affiliation{Jet Propulsion Laboratory, California Institute of Technology, 4800 Oak Grove Drive, Pasadena, CA 91109, USA}

\author[0000-0002-2058-5670]{Karen J.\ Meech}
\affiliation{Institute for Astronomy, University of \Hawaii, 2680 Woodlawn Drive, Honolulu, HI 96822, USA}

\author{Ramon Brasser}
\affiliation{Earth Life Science Institute, Tokyo Institute of Technology, Meguro-ku, Tokyo 152-8550, Japan}

\author[0000-0001-7895-8209]{Marco Micheli}
\affiliation{ESA SSA-NEO Coordination Centre, Largo Galileo Galilei, 1, 00044 Frascati (RM), Italy}

\author{Sukanya Chakrabarti}
\affiliation{School of Physics and Astronomy, Rochester Institute of Technology, 84 Lomb Memorial Dr., Rochester, NY, USA}

\author[0000-0003-0854-745X]{Marc W.\ Buie}
\affiliation{Southwest Research Institute, 1050 Walnut Street, Boulder, CO 80302, USA}

\author[0000-0001-6952-9349]{Olivier R.\ Hainaut}
\affiliation{European Southern Observatory, Karl-Schwarzschild-Stra\ss e 2, 85748 Garching bei M\"unchen, Germany}




\begin{abstract}
The first detected interstellar object \oum\ that passed within 0.25\,au of the Sun on 2017 September 9 was presumably ejected from a stellar system. We use its newly determined non-Keplerian trajectory together with the reconstructed Galactic orbits of 7 million stars from \gdr{2} to identify past close encounters. Such an ``encounter'' could  reveal the home system from which \oum\ was ejected. The closest encounter, at 0.60\,pc (0.53--0.67\,pc, 90\% confidence interval), was with the M2.5 dwarf \object{HIP 3757} at a relative velocity of 24.7\,\kms, 1\,Myr ago. A more distant encounter (1.6\,pc) but with a lower encounter (ejection) velocity of 10.7\,\kms\ was with the G5 dwarf \object{HD 292249}, 3.8\,Myr ago. Two more stars have encounter distances and velocities intermediate to these.
The encounter parameters are similar across six different non-gravitational trajectories for \oum. Ejection of \oum\ by scattering from a giant planet in one of the systems is plausible, but requires a rather unlikely configuration to achieve the high velocities found. A binary star system is more likely to produce the observed velocities. None of the four home candidates have published exoplanets or are known to be binaries. Given that the 7 million stars in \gdr{2} with 6D phase space information is just a small fraction of all stars for which we can eventually reconstruct orbits, it is a priori unlikely that our current search would find \oum's home star system. As \oum\ is expected to pass within 1\,pc of about 20 stars and brown dwarfs every Myr, the plausibility of a home system depends also on an appropriate (low) encounter velocity.
\end{abstract}





\section{Introduction}
\label{sec:introduction}

On 2017 October 19, a fast moving object on an unbound orbit was discovered close to the Earth with the Pan-STARRS1 telescope in \Hawaii.  Now officially named 1I/2017 U1 (\oum), it is the first small body discovered to be unambiguously of interstellar origin and is likely one of many interstellar objects (ISOs) that pass through our solar system \citep{Meech2017}.  Interstellar objects are a natural consequence of planetary formation; they may represent planetesimals that are ejected from a forming planetary system, probably by gravitational kicks from giant planets. Icy planetesimals are expected to be more abundant in the ISO population because the giant planets are relatively inefficient at ejecting small bodies from the dry regions of the inner protoplanetary disk.  Ultimately, a large fraction of the disk mass is likely to be ejected during planet formation and migration \citep{Fernandez1984}, perhaps as much as 1\,M$_{\oplus}$ of planetesimals per stellar mass \citep{Raymond2018a}.  The existence of ISOs like \oum\ can be used to constrain the amount of material ejected during planet formation.  

\oum\ is a small object, with an average radius of 102\,$\pm$\,4\,m assuming a low albedo, and a mass between 10$^9$--10$^{10}$\,kg assuming a comet-like density.  The surface reflectivity is spectrally red as seen by many groups \citep[][and references therein]{Meech2017, Fitzsimmons2018}, suggesting either an organic-rich surface like that of comets and outer solar system asteroids, or a surface containing minerals with nanoscale iron, such as the dark side of Saturn's moon Iapetus.  \oum\ is in a state of excited rotation \citep{Fraser2018, Belton2018, Drahus2018} rotating at a fundamental frequency of 8.67\,$\pm$\,0.34\,h.  The interpretation of its shape from the light curve depends on whether it is in a long-axis or short-axis rotation state; it can either have a cigar-like shape (800\,m in length) or be more of a flattened oval \citep{Belton2018}. There was no visible sign of cometary activity based on upper limits on both the amount of dust that could be present or limits on gas production \citep{Meech2017, Fitzsimmons2018}.   

Newly formed stars tend to have low velocities relative to the Local Standard of Rest (LSR). Due to perturbations from stars, molecular clouds, spiral arms etc., the velocity dispersion of a population increases with time, up to about 50\,\kms\ for stars in the solar neighborhood. Assuming \oum\ was ejected at a low velocity from its home solar system, the low velocity of \oum\ with respect to the LSR suggests it is probably young (although of course a population with a large dispersion still has members with velocities near to zero).\footnote{\label{fn:LSR}The Galactic model we use later, for example, gives the UVW of \oum\ relative to the LSR as ($+0.5$, $-5.6$, $-0.6$)\,\kms, with the LSR moving at 225.1\,\kms\ relative to the Galactic center.}

Astrometric positions of \oum\ were measured from 2017 October 14 (including pre-discovery observations made 5 days before discovery) through 2018 January 2 using both ground-based facilities and HST, with the goal of determining its orbit \citep{Micheli2018}. 
The opportunity to identify the home system for an interstellar object that we characterize up close gives a unique ability to study an exoplanetary system in detail. Our ability to achieve this is limited by the accuracy of its trajectory, by the availability and accuracy of stellar astrometry and radial velocities, and by uncertainties in the Galactic potential, including 
potentially gravitational scattering from stellar encounters.
Various attempts have been make to link the trajectory of \oum\ to a star in the vicinity of the solar neighborhood. These studies propagated \oum's trajectory out of the solar system on the assumption that this was purely gravitational, then integrated its orbit and those of nearby stars to look for close, low-velocity encounters
\citep{Dybczynski2018, Feng2018, Portegies2018, Zhang2018, Zuluaga2018}. 
None of them identified a good candidate.

Since then, a detailed study of \oum's trajectory has shown a 30-$\sigma$ detection of a radial non-gravitational acceleration that can only plausibly be explained by comet-like outgassing \citep{Micheli2018}, albeit from a rather unusual comet.  With this new more accurate information about \oum's trajectory together with the release of the 
larger and more precise \gdr{2} catalog of stellar astrometry and radial velocities, we revisit the experiment of tracing the trajectory back to \oum's home star system.  Our analysis uses a Galactic potential, accommodates the uncertainties in both the stars' 6D phase space coordinates and \oum's trajectory, and uses different models for that trajectory. 

\section{The asymptotic path of \okina Oumuamua toward the Sun}
\label{sec:asymptote}

To compute the asymptotic velocity vector of \oum, we first numerically integrated its trajectory to the year $-3000$, corresponding to a heliocentric distance $28\,000$\,au. Since the motion of {\oum} was not yet rectilinear, we then used a Keplerian extrapolation to derive the asymptotic velocity of {\oum} relative to the solar system barycenter, which we denote as \vinf. Denoting the right ascension and declination of the asymptote 
with (\ra, \dec), the asymptotic velocity vector of {\oum} relative to the solar system barycenter in the International Celestial Reference Frame (ICRF) is
\begin{alignat}{2}
\begin{pmatrix} \vx \\ \vy \\ \vz \end{pmatrix} = 
-\vinf 
\begin{pmatrix} \cos\ra \cos\dec \\ \sin\ra \cos\dec \\ \sin\dec \end{pmatrix} \ .
\label{eqn:vtrans}
\end{alignat}

The detection of non-gravitational motion in the trajectory of {\oum} \citep{Micheli2018} poses additional challenges in estimating its inbound asymptotic motion. In particular, the available astrometric dataset only constrains the detected non-gravitational perturbation from 1\,au to 3\,au on the outbound leg of \oum's trajectory, while its effect on the inbound leg relies on modeling assumptions. To account for the sensitivity of the results to these assumptions, we considered six of the non-gravitational models discussed by \citet[Table 1]{Micheli2018}.  
Table~\ref{tab:asymptote_solutions} gives the different asymptote parameters for the different models.
\begin{table*}
\begin{center}
\caption{Solutions from \citet[Table 1]{Micheli2018} for the asymptotic path of {\oum}. (\ra, \dec) and \vinf\ give the barycentric ICRF direction and asymptotic velocity of {\oum}. These can be transformed to a Cartesian equatorial velocity vector using equation~\ref{eqn:vtrans}. \sigmavx, \sigmavy, \sigmavz\ are the standard deviations in the Cartesian velocity components; the last three columns give the correlation coefficients $\rho$. For reference, we also list JPL solution 13, which is a gravity-only solution that \citet{Mamajek2017} used for an earlier asymptote analysis.
The Galactic coordinates of \candsaa\ are $l=63\degree, b=+17\degree$.
\label{tab:asymptote_solutions}
}
\begin{tabular}{lrrrrrrrrr}
\hline
Solution & \ra   & \dec  & \vinf  & \sigmavx & \sigmavy & \sigmavz & \corvxvy & \corvxvy & \corvxvy \\
      & [deg] & [deg] & [\kms] & [\kms]   & [\kms]   & [\kms]   &          &          &          \\
\hline

\candsaa  & 279.4752 & 33.8595 & 26.420410 & 0.0013481 & 0.0013569 & 0.0000644 &  0.999159 & -0.995578 & -0.993686 \\ 
\candscc  & 279.5247 & 33.8652 & 26.582397 & 0.0012172 & 0.0119929 & 0.0035895 & -0.834714 & -0.533358 &  0.024921 \\ 
\candsbb  & 279.4981 & 33.8633 & 26.402498 & 0.0083704 & 0.0242103 & 0.0121618 &  0.994324 & -0.987668 & -0.998708 \\ 
\candsff  & 279.4987 & 33.8649 & 26.391543 & 0.0101308 & 0.0307732 & 0.0149700 &  0.994705 & -0.989469 & -0.999101 \\ 
\candsdd  & 279.2500 & 33.8185 & 26.572468 & 0.0337819 & 0.1117377 & 0.0560447 &  0.998936 & -0.998156 & -0.999892 \\ 
\candsee  & 279.6067 & 33.8990 & 26.386258 & 0.0050101 & 0.0068670 & 0.0001469 &  0.999764 &  0.530144 &  0.513261 \\ 
JPL 13    & 279.8277 & 34.0020 & 26.320392 & 0.0100903 & 
0.0139452 & 0.0001589 &  0.999956 &  0.897978 & 0.897069 \\
\hline
\end{tabular}
\end{center}
\end{table*}

\begin{figure}
\begin{center}
\includegraphics[width=0.5\textwidth, angle=0]{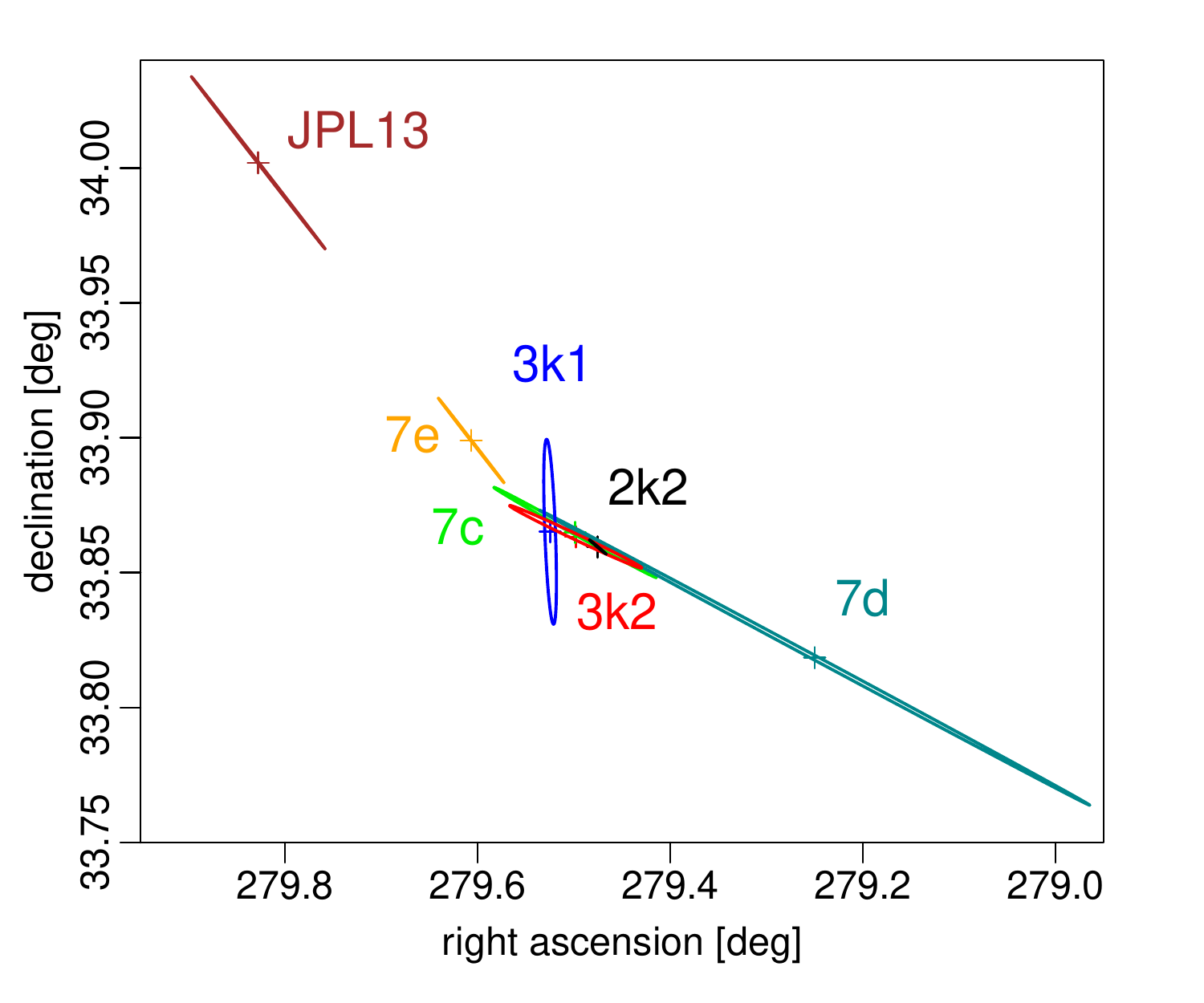}
\caption{Asymptotic right ascension and declination and  error ellipse (90\% confidence region) for six non-gravitational models for \oum's trajectory towards the Sun. For reference, we also show JPL solution 13, which is an earlier gravity-only solution that \citet{Mamajek2017} used for an earlier asymptote analysis.
\label{fig:asymptote_solutions}
}
\end{center}
\end{figure}

We computed the uncertainty of the asymptote estimates through a linear mapping of the covariance of the orbit solution \citep[e.g.][]{Farnocchia2015} for the different non-gravitational models. We ensured the linear approximation was valid by comparing to Monte Carlo simulations. Table~\ref{tab:asymptote_solutions} reports the marginal 1-$\sigma$ uncertainties and correlation coefficients for $(v_x, v_y, v_z)$. Figure~\ref{fig:asymptote_solutions} shows the asymptotes for the different models; the error ellipses encompass 90\% of the probability.
The purely gravitational solution JPL~13 is an earlier solution based on a subset of the data arc and is only for reference; we do not use it because a gravity-only model does not fit the currently available astrometric data \citep{Micheli2018}. 
If we exclude solution 7d, which does not fit the data as well as the other solutions we considered, the scatter is reduced to about $\pm 0.1^\circ$ in both coordinates. The overall uncertainty in the asymptotic velocity, both formal as represented by the error ellipses and systematic because of non-gravitational modeling assumptions, is generally dwarfed by the velocity uncertainties in the candidate home stars. 

\section{Finding close stellar encounters to \okina Oumuamua}
\label{sec:finding}

Our procedure, summarized here, is similar to that used in \cite{2018A&A...616A..37B} to find close stellar encounters with the Sun. Although \oum's path would have to approach within about 0.5\,pc of a star (this being the approximate size of the Sun's Oort cloud) for that to be a plausible origin, due to uncertainties in the Galactic potential
propagating in particular into the estimate of the encounter distance, we consider stars encountering to within about 2\,pc as potentially interesting. Our main source of data is \gdr{2}, but we supplement this with some radial velocities from Simbad (see section~\ref{sec:simbad}).

\subsection{Initial selection}\label{sec:initial_selection}

\gdr{2} provides 6D phase space information -- positions, parallaxes, proper motions, radial velocities -- for 7.2 million stars \citep{2018A&A...616A...1G}. We first select the subset of these that have positive parallaxes, i.e.\ have $\parallax - \parzp > 0$, where $\parzp=-0.029$\,mas is the parallax zeropoint \citep{2018A&A...616A...2L}.\footnote{There is some discussion in the \gaia\ literature about the precise zeropoint for the parallax and even the proper motions, and how they depend on \gmag. For an overview see \url{https://www.cosmos.esa.int/web/gaia/dr2-known-issues}. However, as all home candidates have large values of parallax and proper motion, the exact choice is a minor consideration compared to the larger uncertainties arising from propagating the orbits.} To remove possibly erroneous astrometric solutions, we cut according to the criteria discussed and adopted by \cite{2018A&A...616A..37B}: we remove stars which have {\tt visibility\_periods\_used}\,$ \geq 8$ or unit weight error $u \geq 35$. This leaves us with 7\,039\,430 stars, which we will refer to as the {\em full sample}. The 5th and 95th percentiles on their \gmag-band magnitude distribution are 10.1 and 13.9\,mag.\footnote{Many of the Sun's nearest neighbors are therefore not in the full sample. This includes the alpha Centauri system. Based on the large velocity of \oum\ relative to this (about 37\,\kms), \cite{Mamajek2017} has argued this system is unlikely to be the origin of \oum.}

We then use the linear motion approximation (LMA) to identify which stars in the full sample came within 10\,pc of \oum\ at some time in the past. The LMA, described in \cite{2015A&A...575A..35B} (but modified here to involve the position and velocity of a star relative to \oum), assumes all objects move on unaccelerated orbits. This produces a simple analytic approximation for the three encounter parameters: the time of the encounter relative to 2015.5, \tenc, the separation of the star and \oum\ at encounter, \denc, and their relative velocity at encounter, \venc. We apply the superscript ``lma'' to indicate when these quantities come from the LMA.

For each of the six asymptote solutions for \oum\ in Table~\ref{tab:asymptote_solutions}, we find about 1500 stars which satisfy the conditions $\tenclma<0$ and $\denclma<10$\,pc. 
We call these the {\em encounter candidates}.  As the LMA is not precise, we use this large limit on $\denclma$ to avoid omitting stars which may encounter within 2\,pc when using the orbit integration (the next step). As we summarize later, extending this initial search out to $\denclma<20$\,pc did not uncover additional close encounters.

\subsection{Orbital integration and characterization of encounters}

To improve the accuracy of the encounter parameters, we integrate the paths of the encounter candidates and \oum\ through a smooth Galactic potential.  We characterize the uncertainties at the same time by resampling the initial 6D coordinates of each star (and of \oum) according to their Gaussian covariances, then integrating the orbits of each of the resulting ``surrogate'' objects.
Specifically, for each encounter candidate in each asymptote solution, we draw from the star's 6D Gaussian distribution to give one surrogate star, and then integrate its orbit. We do the same for \oum\ and find the parameters of its encounter with the surrogate star.\footnote{The 3D position of \oum\ is not resampled, because it is -- more or less by definition of its asymptote orbit -- at the solar system barycenter at time $t=0$. This time is the starting point of the integration, the epoch of the Gaia data: J2015.5. Strictly speaking, \oum\ would be offset from this position (in its asymptote orbit) by about 20\,au, but this is completely negligible.}
Repeating this 2000 times produces a 3D distribution over \tenc, \denc, \venc\ for that star and \oum. We characterize the marginal distributions using the median -- \tencmed, \dencmed, \vencmed\ -- as well as their 5th and 95th percentiles (which together form a 90\% confidence interval, CI). This is repeated for each encounter candidate for each of the six asymptote solutions. 

The Galactic potential (a three-component axisymmetric model) and parameters of the Sun's orbit are the same as those described in \cite{2015A&A...575A..35B} and also used in \cite{2018A&A...616A..37B}. Encounter parameters depend only weakly on these choices because it is only {\em differences} in the potential experienced by the stars and \oum\ which influence their mutual encounters. Note that the velocity of \oum\ relative to the LSR mentioned in section~\ref{sec:introduction} is a consequence of this model, not an input. We neglect the mutual gravity between individual stars and between individual stars and \oum\ (the latter at least is probably irrelevant; see section~\ref{sec:discussion}). 

\section{Encounter results}
\label{sec:encounter_results}

The set of close encountering stars -- those with $\dencmed<2$\,pc -- is rather similar among the six asymptote solutions. In all cases, the five closest encounters always come from a single set of six stars.
We therefore present detailed results for one baseline solution, \candsaa, then mention results for other solutions as appropriate.

\begin{figure}
\begin{center}
\includegraphics[width=0.5\textwidth, angle=0]{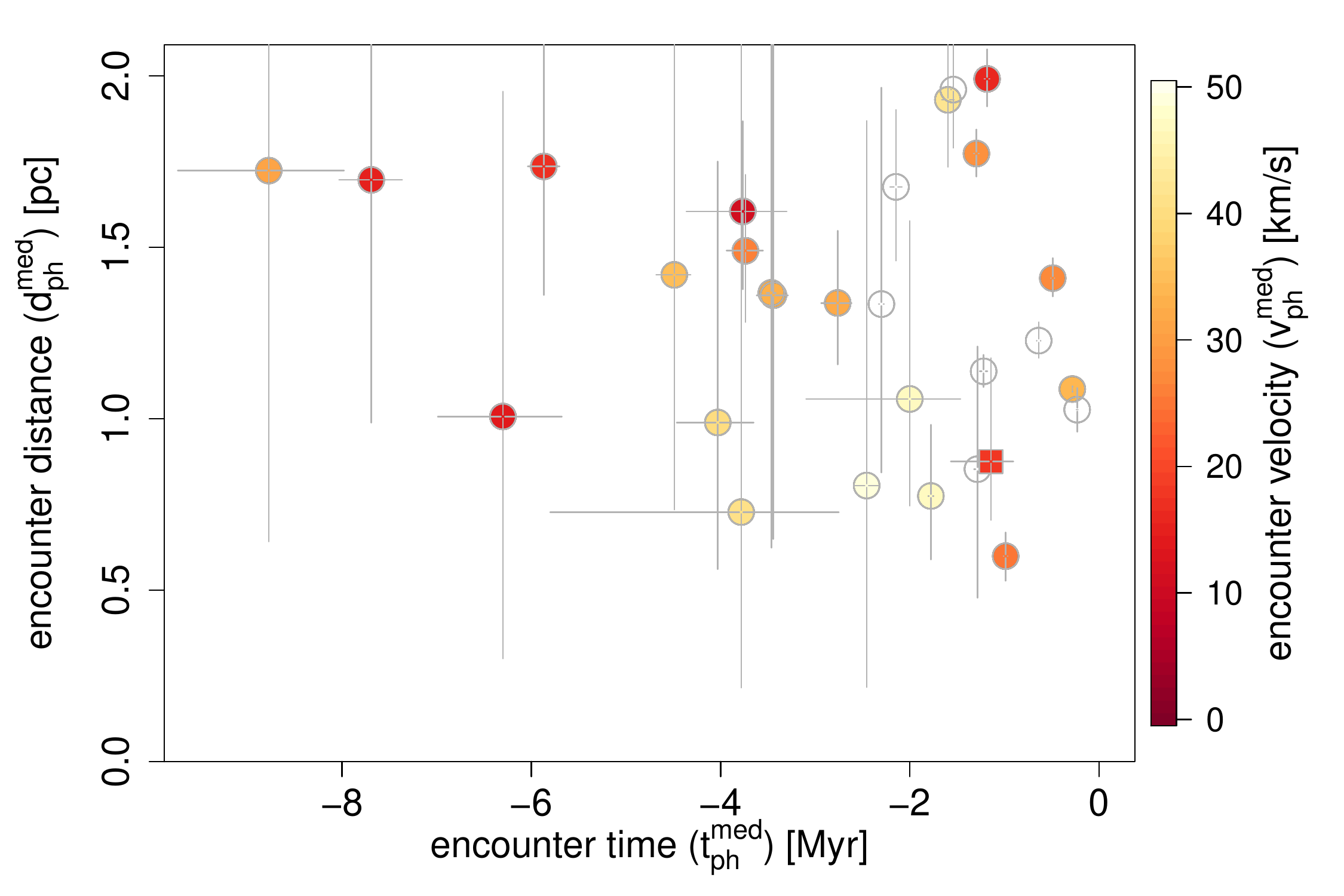}
\caption{Encounter parameters for the 28 stars from \gdr{2} that encounter \oum\ (solution \candsaa) with \dencmed$<2$\,pc, plus \object{home-4} (the square symbol). The point shows the median; the error bars show the extent of the 5th and 95th percentiles (the error bars are smaller than the points in some cases). The smallest encounter velocity plotted is 10.7\,\kms\ (\object{HD 292249}). White circles have $\vencmed>50$\,\kms.
\label{fig:dph_vs_tph_withCIs_colour_is_vph}
}
\end{center}
\end{figure}

\begin{sidewaystable*}
\centering
\caption{The 28 stars from \gdr{2} that encounter \oum\ (solution \candsaa) in the past with a median encounter distance (\dencmed) below 2\,pc, sorted by this value (plus \object{home-4}, listed at the bottom). Columns 2, 5, and 8 are \tencmed, \dencmed, and \vencmed\ respectively. The columns labeled 5\% and 95\% are the bounds of the corresponding confidence intervals. Columns 11--16 list the parallax ($\parallax-\parzp$), total proper motion (\propm), and radial velocity (\vr) along with their 1-sigma uncertainties.  Columns 17, 18, and 19 are the apparent and absolute \gaia\ G-band magnitude (assuming zero extinction) and \gaia\ color respectively. $^{\dag}$\,=\,\object{HIP 3757 } $^{\ddag}$\,=\,\object{HD 292249} $^\P$\,=\,\object{home-3}, $^{\sharp}$\,=\,\object{home-4}.
\label{tab:periStats}
}
\tabcolsep=0.14cm
\begin{tabular}{*{19}{r}}
\hline
1 & 2 & 3 & 4 & 5 & 6 & 7 & 8 & 9 & 10 & 11 & 12 & 13 & 14 & 15 & 16 & 17 & 18 & 19 \\
\hline
\gdr{2} source ID & \multicolumn{3}{c}{\tenc [kyr]} &  \multicolumn{3}{c}{\denc [pc]} &  \multicolumn{3}{c}{\venc [\kms]} &
 \parallax & $\sigma(\parallax)$ & \propm & $\sigma(\propm)$ & \vr & $\sigma(\vr)$ & \gmag & \mg & \bprp \\
         & med & 5\% & 95\% & med & 5\% & 95\% & med & 5\% & 95\% & \multicolumn{2}{c}{mas} & \multicolumn{2}{c}{\maspyr} &  \multicolumn{2}{c}{\kms} & mag & mag & mag \\
\hline
\input{figures/cands41plushome4_encounters.tex}
\hline
\end{tabular}
\end{sidewaystable*}

\begin{figure*}
\begin{center}
\includegraphics[width=\textwidth, angle=0]{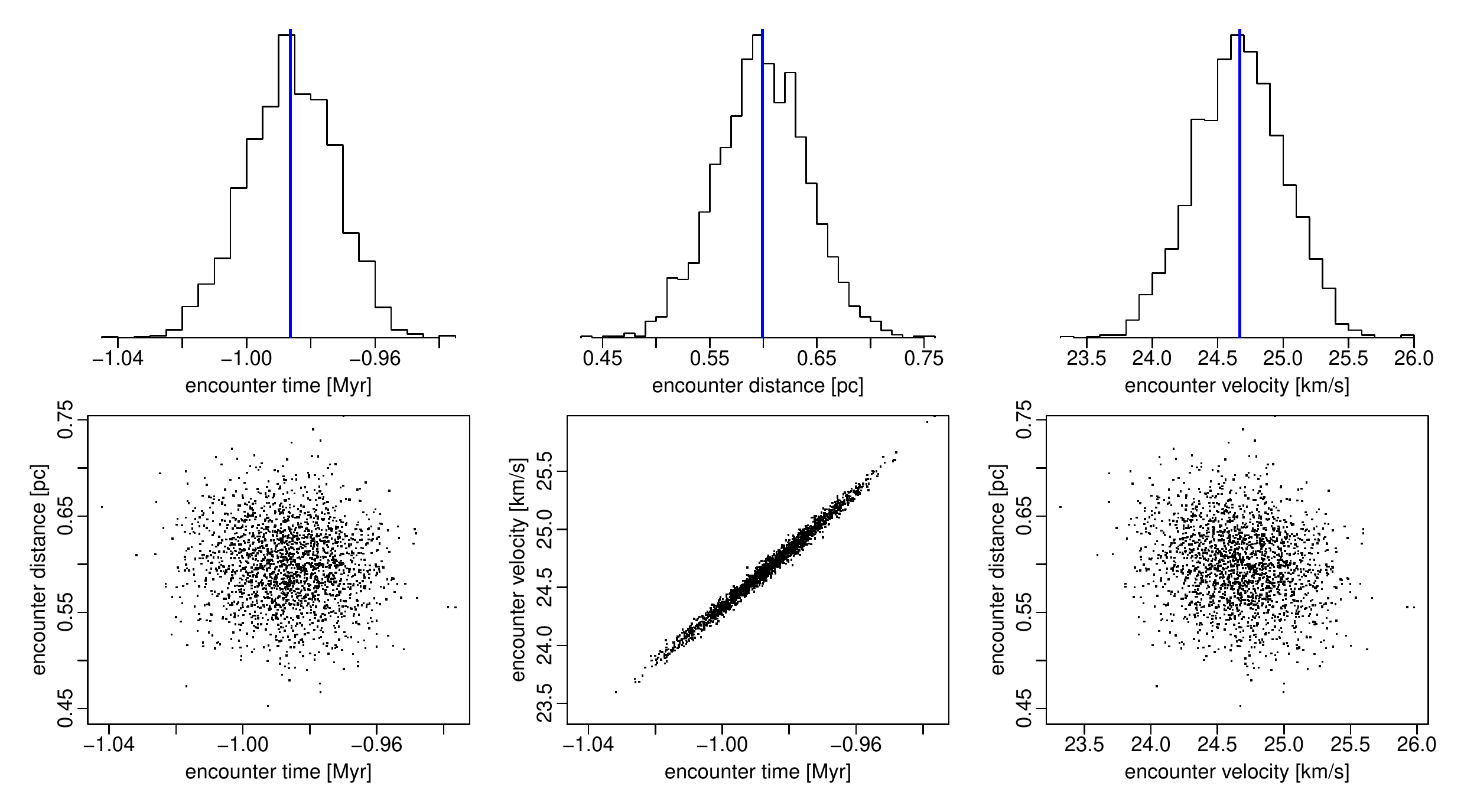}
\caption{Distribution over encounter parameters (for the 2000 surrogates) of \oum\ asymptote solution \candsaa\ with \object{\gdr{2} 2525688198820543360} (= \object{HIP 3757}). The vertical blue line shows the median of the distribution.
\label{fig:cands41_2525688198820543360_perisamp}
}
\end{center}
\end{figure*}

For solution \candsaa, 28 stars have $\dencmed<2$\,pc. Figure~\ref{fig:dph_vs_tph_withCIs_colour_is_vph} plots encounter distance versus encounter time, with the encounter velocity shown in color (the scale intentionally saturates at 50\,\kms). Table~\ref{tab:periStats} gives their details. 
The closest encounter is \object{\gdr{2} 2525688198820543360} (= \object{HIP 3757}), with $\tencmed=-0.99$\,Myr, $\dencmed=0.60$\,pc, and $\vencmed=24.7$\,\kms.  The distributions over the surrogates for this star (Figure~\ref{fig:cands41_2525688198820543360_perisamp}) show that the encounter parameters are reasonably well constrained. The strong correlation between encounter time and velocity is typical for such encounters (as there is usually a component of the relative velocity along \oum's path, so a change in velocity corresponds to a change in time to encounter). \object{HIP 3757} is a high proper motion 
M2.5 dwarf \citep{2003AJ....126.2048G} currently 25\,pc from the Sun.   Its absolute magnitude and color 
place it on the main sequence (see Figure~\ref{fig:cmd} below; compare to Figure 8 of  \citealt{2018A&A...616A..37B}). Based on the same data, \gdr{2} lists its \teff\ as 3950\,K (68\% CI 3790--4010\,K) with negligible extinction \citep{2018A&A...616A...8A}. RAVE-DR5 \citep{2017AJ....153...75K} gives the \teff\ as $3680 \pm 150$\,K and metallicity as $-1.7 \pm 0.6$\,dex ($\log\rm{g}$ is poorly constrained).

The slowest encounter for solution \candsaa\ is \object{\gdr{2} 4899996487129314688}, with $\tencmed=-5.2$\,Myr, $\dencmed=3.6$\,pc and $\vencmed=6.5$\,\kms. Although this low velocity is more compatible with possible ejection scenarios (see section~\ref{sec:ejection_mechanisms}), the large encounter distance -- the 90\% CI is 3.11--4.16\,pc -- rules this out as an origin.

Considering median encounter distances up to 2\,pc,
the slowest encounter is \object{\gdr{2} 3107000885484340224} (= \object{HD 292249}) with $\tencmed=-3.8$\,Myr, $\dencmed=1.60$\,pc, and $\vencmed=10.7$\,\kms. This is a G5 star (according to the Henry Draper Catalog as presented in \citealt{1995A&AS..110..367N}). \gdr{2} places it on the main sequence with $\teff=5040$ (90\% CI 5000--5100)\,K.

\begin{figure}
\begin{center}
\includegraphics[width=0.5\textwidth, angle=0]{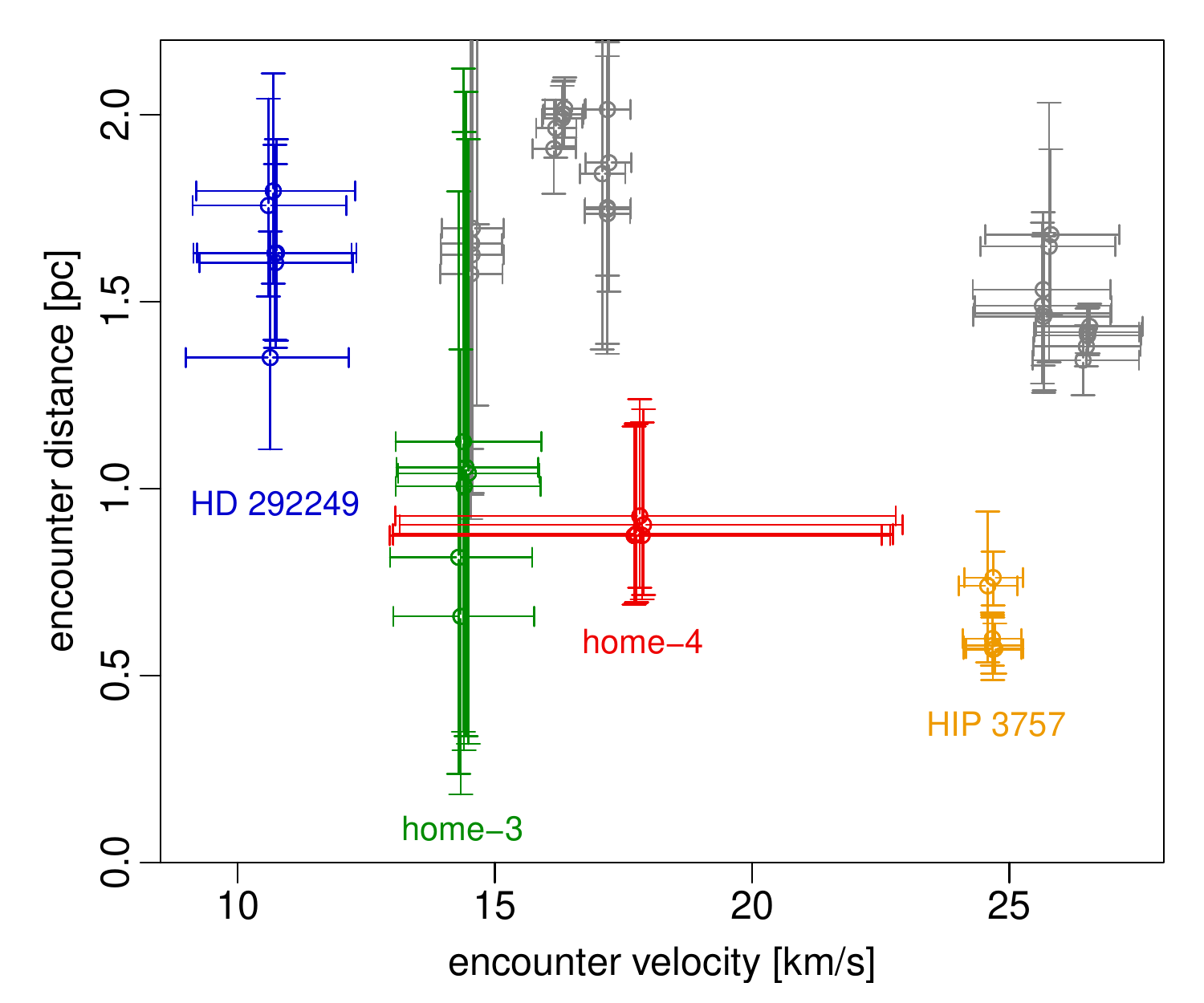}
\caption{Encounter distance vs.\ velocity between \oum\ and the four main home candidate stars -- 
\object{\gdr{2} 2525688198820543360} (= \object{HIP 3757}) in orange, \object{\gdr{2} 3107000885484340224} (= \object{HD 292249}) in blue, \object{\gdr{2} 2502921019565490176} (= \object{home-3}) in green, 
\object{\gdr{2} 3666992950762141312} (= \object{home-4}) in red,
 -- for all six asymptote solutions.
The circles show the median of the encounter distributions; the error bars span the 90\% confidence interval. From closest to most distant (in the median) the solutions are
\candsff, \candsee, \candsbb, \candsaa, \candsdd, \candscc\
for \object{HIP 3757},
\candsdd, \candsaa, \candsbb, \candsff, \candscc, \candsee\
for \object{HD 292249}, and
\candscc, \candsdd, \candsaa, \candsee, \candsbb, \candsff\
for \object{home-3}.
Also shown (in gray) are the other five home candidates 
that have $\dencmed<2$\,pc and $\vencmed<28$\,\kms\ in \candsaa. They are, from left to right, 
\object{\gdr{2} 2264912868532666240},
\object{\gdr{2} 3234412606443085824},
\object{\gdr{2} 2425971152835667072}, 
\object{\gdr{2} 4319754583064325248}, and
\object{\gdr{2}  211810233512673792}.
\label{fig:dph_vs_vph_multistars}
}
\end{center}
\end{figure}

Looking at the orbital integration results for all six solutions for \oum, we find that \object{HIP 3757} is the closest encounter in four cases, and \object{HD 292249} is the slowest (for $\dencmed<2$\,pc) in all six cases. 
The encounter velocities and distances for these six solutions are shown in Figure~\ref{fig:dph_vs_vph_multistars}.  The relatively small spread across the six solutions, especially in velocity, is a result of the small spread across the asymptote solutions compared to the uncertainties in the 6D coordinates of the stars.

The closest encounter for solution \candscc\ is \object{\gdr{2} 2502921019565490176}, where it has \dencmed=0.66\,pc (90\% CI 0.18--1.37\,pc) and \vencmed=14.3\,\kms (90\% CI 13.0--15.8\,\kms). This is somewhat closer than its approach in solution \candsaa\ shown in Table~\ref{tab:periStats}, where it is the seventh closest encounter at \dencmed=1.01\,pc. Note the relatively large uncertainty in its encounter distance (for all solutions), a consequence of its particular astrometric and radial velocity data.
This star has a good compromise between encounter proximity and velocity, as can seen in Figure~\ref{fig:dph_vs_vph_multistars} (plotted in green). This star has no entry in Simbad. It is clearly visible in DSS, SDSS, WISE, and 2MASS images, and based on proximity and magnitude it is probably \object{2MASS J02335086+0144054}. We call it ``home-3'' for short. The \gdr{2} catalog assigns it \teff\,=\,4320 (68\% CI 4230--4470)\,K.

The closest encounter for solution \candsdd\ is \object{\gdr{2} 3174397001191766400}, which has similar encounter parameters as in Table~\ref{tab:periStats} (where it is the second closest encounter). The high encounter velocity for this star (40.9\,\kms) makes it an unlikely home candidate. The large uncertainty in its encounter velocity -- the 90\% CI is 25.8--55.9\,\kms\ -- 
is a direct result of the large uncertainty in its radial velocity, which is likely a consequence of its low signal-to-noise ratio RVS spectrum.


\begin{figure}
\begin{center}
\includegraphics[width=0.5\textwidth, angle=0]{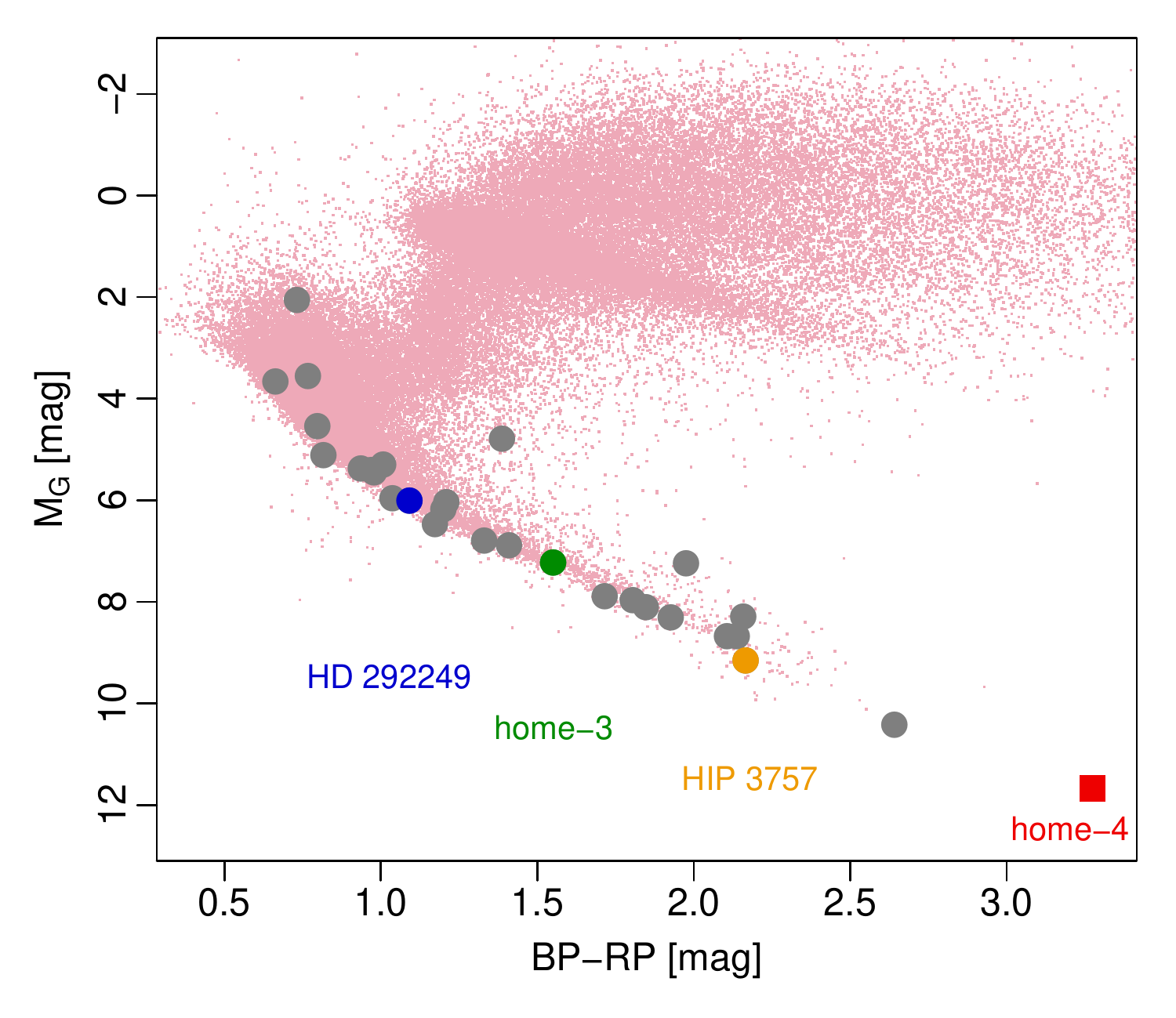}
\caption{The color absolute-magnitude diagram (assuming zero extinction, with distances from \citealt{2018AJ....156...58B}) 
for 1\% of the full sample as small pink points and the 29 close encounters from Table~\ref{tab:periStats} as large gray points.
The four main home candidates are colored and labeled.
\label{fig:cmd}
}
\end{center}
\end{figure}

Figure~\ref{fig:cmd} shows the color absolute-magnitude diagram (CMD) of the close encounters from Table~\ref{tab:periStats} compared to the full sample. Most of the encountering stars are main sequence stars, and we see that they -- in particular the fainter, lower mass main sequence stars -- are overrepresented relative to the full sample. 
This is the result of a volume effect, arising because (a) these stars are intrinsically faint, and so have to be near to the Sun to be in the magnitude-limited full sample, and (b) stars which are currently close to the Sun are far more likely to have come close to a trajectory leading toward the Sun in the recent past (i.e.\ \oum's) than are more distant stars. 
This also explains the preponderance of close encounters within the past few Myr (see 
\citealt{2018A&A...616A..37B} for further discussion).

\subsection{A fourth home candidate found by supplementing \gdr{2} with Simbad radial velocities}\label{sec:simbad}

As \gdr{2} does not list radial velocities for most of its 1.33 billion sources that have astrometry, we searched Simbad for non-\gaia\ radial velocities for these sources. We found 222\,275 stars which also satisfy the criteria on parallax and parallax metrics in section~\ref{sec:initial_selection}. Of these, 476
have $\tenclma<0$ and $\denclma<10$\,pc for solution \candsaa\ (and similar for the other solutions). After performing the orbital integration, 13 have $\dencmed<2$\,pc.

The closest encounter with \oum\ is at 0.7\,pc, but with a relative velocity of 266\,\kms, so is uninteresting.
The only good home candidate is the second closest encounter in the sample, \object{\gdr{2} 3666992950762141312}, which we name ``\object{home-4}''.
Its closest approach to \oum\ of 0.9\,pc was 1.1\,Myr ago with a relative velocity of 18\,\kms. Full details are given in the last line of Table~\ref{tab:periStats}, and it is plotted in Figures~\ref{fig:dph_vs_vph_multistars} and~\ref{fig:cmd} in red. The relatively large uncertainty in its encounter velocity is mostly due to the low signal-to-noise of its radial velocity (which is from \citealt{2015ApJS..220...16T}). \cite{2015A&A...577A.128A} give its spectral type as M5V, which agrees with its position in the CMD (Figure~\ref{fig:cmd}) and the \teff\ of 3530 (68\% CI 3290--4100)\,K listed in \gdr{2}. It has several names and references in Simbad, and appears in various catalogues of nearby M dwarfs.

The other 11 encounters are more distant and faster, and are less interesting than other candidates already mentioned.

\subsection{Summary of main home candidates}

A few other stars in Table~\ref{tab:periStats} have close approaches to \oum, but in all cases their velocities are rather too high for plausible ejection scenarios, and/or their encounter parameters have large uncertainties. We therefore consider \object{HIP 3757},
\object{HD 292249},
\object{home-3}, and \object{home-4} as our main home candidates.  All four stars have good astrometric and radial velocity measurements and quality metrics in \gdr{2}. \object{HIP 3757}'s astrometry is consistent with that in both Hipparcos \citep{1997A&A...323L..49P} and Hipparcos-2 \citep{2007ASSL..350.....V} (it is not in TGAS). \object{HD 292249}'s astrometry agrees with that in TGAS \citep{2016A&A...595A...2G} (it is not in Hipparcos). 
None of these stars are known to harbor exoplanets or to be binary systems.

To ensure that our LMA-based selection with $\denclma<10$\,pc was not too restrictive, we extended it to $\denclma<20$\,pc. This produced a further 4500 encounter candidates (for the complete \gaia\ data), for which we ran the orbital integration with surrogates. Only two of these candidates had $\dencmed<2$\,pc, both at around 1.7\,pc, but with large 90\% CIs (0.4--4.9\,pc) and velocities of 25\,\kms\ and 47\,\kms.

\subsection{Future encounters}

We can use the same procedure described so far to determine \oum's {\em outbound} trajectory from the Sun and identify {\em future} close encounters. For model \candsaa\ in \cite{Micheli2018}, the asymptote is $(\ra, \dec, \vinf)=(357.8512\degree, 24.7164\degree, 26.40776\,\kms)$ (Galactic coordinates $l=106\degree, b=-36\degree$). We searched both the \gaia\ sample and the Simbad supplement. There are not as many close/slow encounters as for the inbound trajectory. 

The closest encounter is with the high proper motion star \object{\gdr{2} 2279795617408467712} (= \object{TYC 4600-1769-1}) at a distance of 0.332 (90\% CI 0.313-0.351)\,pc, which takes places in 0.716\,Myr with a large relative velocity of 112\,\kms. 
One of the slowest encounters will be with \object{\gdr{2} 335659463780716288} in 2.7\,Myr with median velocity of 30\,\kms\ and distance of 1.4\,pc.
In just 0.029\,Myr \oum\ will pass the M5 dwarf \object{Ross 248} (=\object{\gdr{2} 1926461164913660160}) at 0.459 (90\% CI 0.458--0.460)\,pc with a relative velocity of 104\,\kms. \object{Ross 248} is currently one of the closest stars to the Sun, at 3.15\,pc.

\section{Ejection mechanisms}
\label{sec:ejection_mechanisms}

Several mechanisms have been proposed in the literature for ejecting comets or asteroids from planetary systems. Most of these rely on studies of the formation of the Oort cloud around the Sun, e.g. \cite{Fernandez1984,Duncan87,T93,BM13}. Here we discuss scattering by a giant planet and a very close stellar passage.

\subsection{Scattering by a giant planet or a binary star}

We consider a system comprising a central star with mass $M_*$, a giant planet with mass $m_{\rm p}$ on a (nearly) circular orbit of semi-major axis $a_{\rm p}$, and planetesimals such as \oum. Numerical simulations of Oort cloud formation indicate that planetesimals usually undergo a random walk through diffusion up to escape energies. This diffusion occurs through distant encounters with the planet(s) in which the periastron distance is $q\sim\frac{9}{8}a_{\rm p}$ \citep{Duncan87,BM13}. This evolution eventually leads to highly eccentric orbits, because the energy changes occur at periastron and the semi-major axis undergoes a random walk while the periastron distance remains approximately constant \citep{Duncan87}. If we describe the total specific energy of a planetesimal such as \oum\ by $-1/(2a)$
then the typical energy change during each periastron passage due to distant interactions with the planet is \citep{Duncan87,T93}

\begin{equation}
\vert\Delta E\vert = \frac{10}{a_{\rm p}}\frac{m_{\rm p}}{M_*}\: {\rm au}^{-1},
\end{equation}
which is applicable for low-inclination orbits; at higher inclination the energy kicks are gentler. Before its last periastron passage, \oum\ is expected to have a total specific energy of $-\frac{1}{2}\Delta E$. After its last periastron passage it will typically have a total energy $\frac{1}{2}\Delta E$, and the typical escape velocity from the solar system is then
\begin{equation}
v_{\rm esc}=\sqrt{\frac{GM_*\Delta E}{2}} \ .
\label{eqn:vesc}
\end{equation}
For encounters with Jupiter $\Delta E = 2 \times 10^{-3}$ au$^{-1}$ and $v_{\rm esc} \sim 1\,\kms$; these values are consistent with various numerical studies \citep{Duncan87,BM13}. In this simple approximation, $v_{\rm esc} \propto (m_{\rm p}/a_{\rm p})^{1/2}$, so that a Jupiter-mass planet at 1\,au will eject planetesimals with a typical velocity close to 2\,\kms. Ejection with a velocity close to the inferred encounter velocity of 10\,\kms\ or more from section~\ref{sec:encounter_results} would require either an extremely massive object close to the central star (such as a binary system), or a rare, deep encounter with a giant planet, for which $v_{\rm esc} \sim v_{\rm p}$, the orbital velocity of the planet ($\sim$ 13\,\kms\, for Jupiter).

Ejection velocities much higher than $v_{\rm p}$ are rare because they require that after the encounter, the velocity vector of the planetesimal be parallel to that of the planet. Since encounters are isotropic, the volume of outcomes that achieve this is small. Immediately after such an encounter, the orbital velocity of the planetesimal is $v_{\rm tp}+v_{\rm p}$ and ejection is only possible when $v_{\rm tp}+v_{\rm p} > \sqrt{2}v_{\rm p}$, where $v_{\rm tp}$ is the orbital velocity of the planetesimal before the encounter. Ejection in this manner therefore requires $v_{\rm tp} \geq (\sqrt{2}-1)v_{\rm p}$. Such a high encounter velocity, and thus ejection velocity, are only possible if the planetesimal crosses the  planet's orbit. Numerical simulations show, however, that most planetesimals are ejected on orbits with perihelia greater than the planet's orbit \citep{BD08}.

\begin{figure}
\begin{center}
\includegraphics[width=0.5\textwidth, angle=0]{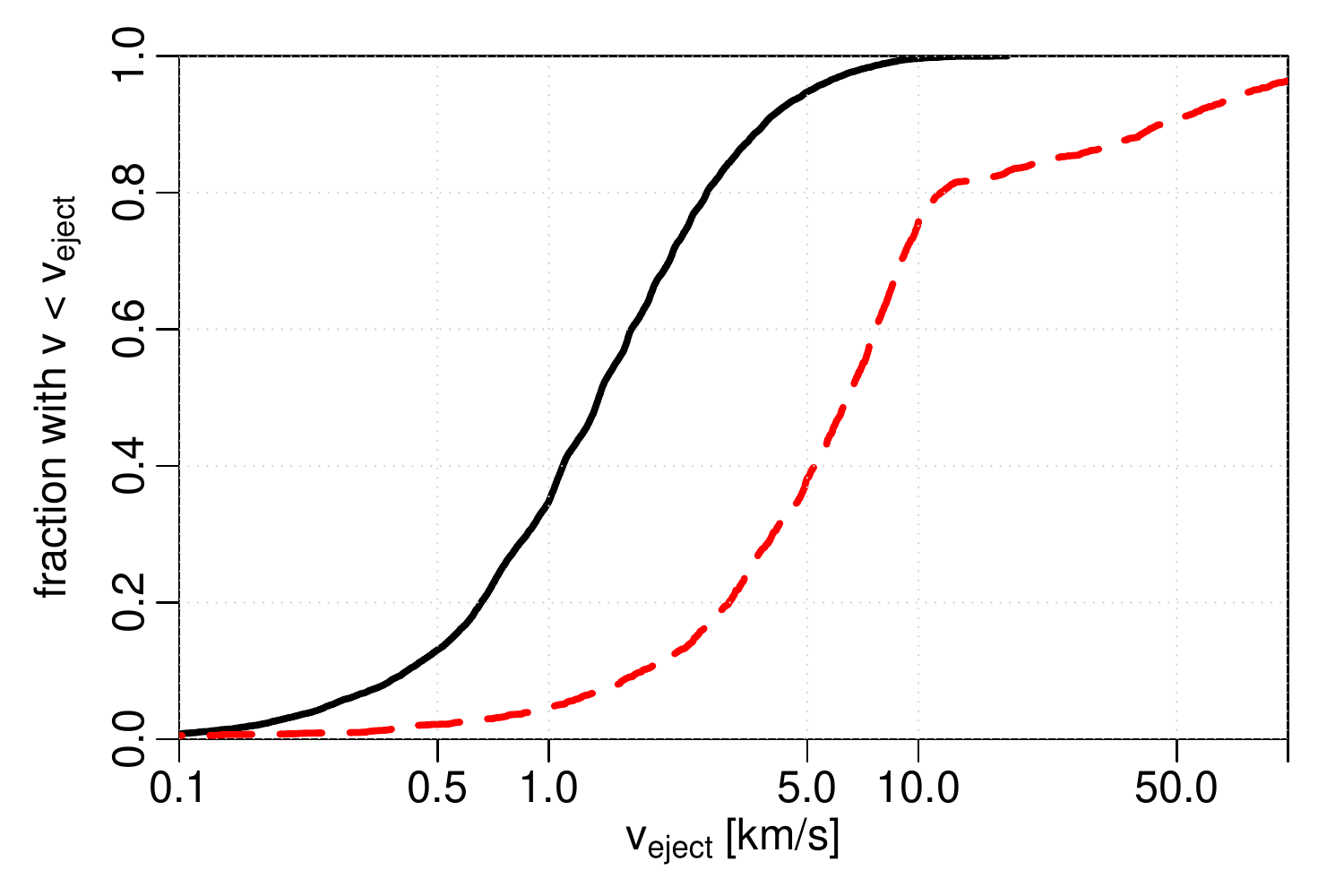}
\caption{Cumulative distribution of the ejection velocity of planetesimals scattered by (i) the giant planets in the solar system (black line; data from \citealt{BM13}), and (ii) a binary system with stars of masses 0.1\,\msun\ and 1.0\,\msun\ separated by 10\,au in a mutual circular orbit (dashed red line).
The maximum velocity for (i) is 17.4\,\kms\ (note the log scale).}
\label{fig:veject}
\end{center}
\end{figure}

Figure~\ref{fig:veject} shows the cumulative distribution over the ejection velocity for planetesimals that are scattered by the giant planets in the solar system. The duration of the simulation is 200\,Myr, wherein over 90\% of planetesimals were removed. Since the giant planets are assumed to be static, the distribution of ejection velocity will reach a steady state on a timescale that is comparable to the dynamical evolution timescale of the planetesimals. For encounters with Jupiter and Saturn this is about 3\,Myr, while for encounters with Uranus and Neptune it approaches 100\,Myr \citep{Duncan87}. The  majority ($>$70\%) of the planetesimals are ejected by Jupiter \citep{BD08}, so that the distribution is expected to be close to that from ejection solely by Jupiter. The median ejection velocity of 1.3\,\kms\ corresponds closely to the value derived from equation~\ref{eqn:vesc}. The cumulative distribution of the ejection velocity is approximately log-normal, with a mean of 0.13 and a standard deviation of 0.37. 
An ejection velocity of 10\,\kms\ or greater occurs only in 0.3\% of the cases.
For gas giants residing closer to their host star, the distribution is expected to look similar but shifted towards higher velocities, scaling as $a_{\rm p}^{-1/2}$. A Jupiter-mass planet at 1\,au, therefore, is expected to have a 5\% probability of ejecting planetesimals at 10\,\kms, because the curve will be shifted to the right by a factor of roughly two and the 95$^{\rm th}$ percentile for the solar system is then at 5.1\,\kms. The probability that such a planet ejects \oum\ at 20\,\kms\ or greater is still only 0.3\%.

\cite{Raymond2018a} note that a planetary system with initially unstable gas giants is a more efficient source of interstellar planetesimals than a system with stable giant planets. The reason is that during a phase of instability, multiple gas giant planets will be on highly eccentric orbits and can thus eject planetesimals over a much wider range of orbital radii than a stable system \citep{Raymond2011,Raymond2012}, although it is unclear from the simulations whether such highly eccentric planetary systems eject planetesimals at higher velocities.

Higher ejection velocities can occur for planetesimals scattered in a binary star system. To demonstrate this, we performed a simple dynamical experiment on a system comprising a 0.1\,\msun\ star in a 10\,au circular orbit about a 1.0\,\msun\ star. (This is just an illustration; a full parameter study is beyond the scope of this work.) Planetesimals were randomly placed between 3\,au and 20\,au from the primary, enveloping the orbit of the secondary. The cumulative distribution of the ejection velocities is shown as the red dashed curve in Figure~\ref{fig:veject}.
Once again most (80\%) of the ejections occur at velocities lower than 10\,\kms, but a small fraction is ejected at higher velocities in the range of those we observe (and even exceeding 100\,\kms).
We mention here that the second closest encounter in Table~\ref{tab:periStats}, 
\object{\gdr{2} 3174397001191766400}, lies above the main sequence in Figure~\ref{fig:cmd}, suggesting it might be a binary.

\subsection{Interaction with a second star}

A star passing close to a home candidate star could in principle eject small bodies such as \oum. 
The more massive/closer/slower the transit, the more likely the ejection. 
If we assume that \oum\ was originally residing on a wide, weakly-bound orbit around a solar-type star, then the change in velocity imparted on it by the impulse of a passing star is \citep{Fouchard11}
\begin{equation}
\Delta V = \frac{2GM_*}{v_*b}
\label{eqn:deltav}
\end{equation}
where $b$ is the impact parameter (approximately equal to the closest distance if $v_*$ is high). 
(This approximation does not apply if either star is a binary system.)
A passing star with $M_*=0.5$\msun\ and $v_* = 25\,\kms$ requires an encounter within 35\,au to impart $\Delta V > 1\,\kms$.
Thus it is a priori very unlikely that \oum\ would be ejected by a typical passing star. 

We nonetheless ran models to determine whether any of the 7 million stars in the full sample 
passed close to the home candidates at the right time, e.g.\ within the 90\% CI on \tenc\ given in Table~\ref{tab:periStats}. 
No good encounters were found for the first three home candidates, \object{HIP 3757}, \object{HD 292249}, and \object{home-3}. Two encounters with \object{home-4} from the LMA were promising enough to be further investigated further with the orbital integration. One of these, \object{\gdr{2} 2984656203733095936}, had a median closest approach of 1.4\,pc at $-1.25$\,Myr (the 90\% CI is $-1.38$ to $-1.14$\,Myr, so overlaps considerably with \object{home-4}'s encounter with \oum) and median encounter velocity of 42\,\kms. More significantly, it has a 20\% chance of approaching closer than 0.51\,pc to \object{home-4}. This star is in fact listed in Table~\ref{tab:periStats} (not that surprising, as all three objects must come close together at the same time). Its absolute magnitude and color, together with its \teff\ of 3960 (68\% CI 3870--4010)\,K from \gdr{2}, imply it is a late K or early M dwarf, so not very massive. Equation~\ref{eqn:deltav} suggests it is extremely unlikely to have ejected \oum\ (as $b\sim 1$\,au when $v_*=42\,\kms$, $M_*=0.5$\msun, and $\Delta V=17.9\,\kms$).
The other star, \object{\gdr{2} 4294690189048270464}, also has an encounter in roughly the right time range (90\% CI $-0.96$ to $-0.90$\,Myr) but has a large encounter distance spread (90\% CI 0.52--4.97\,pc), so is even less favorable.


In conclusion, we find no convincing star which could have ejected \oum\ from a home candidate star. There could, of course, be other candidates for which we do not have 6D data.

\section{Discussion}
\label{sec:discussion}

\subsection{Limiting assumptions}

{\em Modeled motions.} The derivation of the \gdr{2} astrometry assumes that stars move on unaccelerated paths over the time range of \gaia's observations. In particular, it assumes that any binarity does not affect the astrometry of their observed center-of-light. We have also assumed that the non-gravitational forces inferred for \oum's observed outbound trajectory apply also to its inbound trajectory.

{\em Restricted sample.} The most obvious limitation of our work is that it is restricted to just those 7 million stars in \gdr{2} with 6D phase space information. This is a small fraction of the stars within some distance horizon
for which one could in principle look for home candidates, if the data were available. There are, for example, 35 million stars in \gdr{2} within 500\,pc \citep[most probable distances from][]{2018AJ....156...58B}, fives times as many as the full sample. For comparison, the median distance of the full sample is 1250\,pc; 180 million stars in \gdr{2} are closer than this.
Hence it is a priori unlikely that we would find the home system in our study. Even when we have full data, searching out to larger distances (100s of pc, and thus \oum\ travel times of order 10\,Myr or more) will be difficult, as uncertainties in the Galactic potential become more relevant and N-body effects eventually become significant \citep[e.g.][]{Zhang2018}.

{\em Low probability encounters.} Table~\ref{tab:periStats} lists all the stars which have $\dencmed <2$\,pc and therefore have a probability greater than 0.5 of passing within 2\,pc of \oum. There are, of course, stars with $\dencmed>2$\,pc which could have a non-negligible probability of encountering closer than 2\,pc, and thus are potentially interesting. For solution \candsaa\ there are 1440 stars with $2 \leq \dencmed \leq 10$\,pc. 66 of these have a probability greater than 0.01 of encountering closer than 2\,pc. Their median probability is only 0.05, however.
In comparison, the 28 candidates in Table~\ref{tab:periStats} have a median probability of 0.987 of encounter closer than 2\,pc, and the individual home candidates have probabilities of 1.000 (\object{Hip 3757}), 0.989 (\object{HD 292249}), and 0.957 (\object{home-3}). 
None of these 66 stars have encounter distance distributions which peak below 1.5\,pc, so none are close encounters even in a maximum likelihood sense.

{\em Close encounters inevitable.} With the same full sample as used here, \cite{2018A&A...616A..37B} inferred the rate of stellar encounters (all masses) within 1\,pc of the Sun to be $19.7 \pm 2.2$ per Myr. The completeness of the survey was estimated to be about 0.15, so one would expect to see three encounters within 1\,pc per Myr in \gdr{2}. To a first approximation this rate should also be valid for \oum. Thus when tracing \oum's orbit over a few million years, it is not at all surprising that we do find a few close encounters. 
Of course, the closer the encounter, the more plausible it is as the home star. But given the inevitable imperfection of the model Galaxy potential, plus the fact that the inferred encounter separation is more sensitive to this than the inferred encounter velocity, a low encounter velocity is an important criterion for identifying a home candidate.

{\em Smooth potential.} Our orbital integration uses a smooth potential, thereby neglecting N-body interactions between \oum\ and individual stars. A simple calculation of the deflection by passing stars (see section 5.3 of \citealt{2015A&A...575A..35B}) shows, however, that these can be neglected for more recent encounters. 
For example, \oum\ travels about 26\,pc on its (time-reversed) journey from the Sun to encounter \object{HIP 3757} about 1\,Myr ago. (Recall that
$1 \kms \simeq 1$\,pc\,Myr$^{-1}$.) \oum\ would have to come within 0.008\,pc of a 1\,\msun\ star during this journey in order for its closest approach at \object{HIP 3757} to be deflected by 0.05\,pc or more (an amount that is well below the spread in \denc\ arising from the uncertainties in the 6D coordinates). 
Adopting a conservatively large local stellar number density of 1\,pc$^{-3}$, the probability that \oum\ would have one or more such close encounters on this journey is just 0.005. 
Potentially more significant is gravitational interactions between stars deflecting each other's paths, and thus changing their encounters with \oum.
As the referee emphasized, this becomes more relevant if \oum\ was ejected at a low velocity from its home star (as favored by the giant planet scattering mechanism discussed in section~\ref{sec:ejection_mechanisms}), as then the home star too has traveled a long distance. For example, if \oum\ were ejected at 1\,\kms\ 10\,Myr ago, this would have occurred 260\,pc from the Sun, yet the home star would currently be only 10\,pc away from \oum\ and hence from the Sun. Tracing back its orbit in the absence of relevant perturbations over a long path means we could fail to unite it with \oum\ within the separation limit adopted.

{\em N-body effects.} \cite{Dybczynski2018} took into account the gravitational interactions of 57 stars in their study of the possible origin of \oum\ (discussed in the next section), although they neglected the (more significant) uncertainties in the stellar astrometry and radial velocities. 
We explicitly tested the effects of including the gravity of the closest 100 closest encounters to \oum\ (identified from the orbital integration with solution \candsaa), on the orbits of both \oum\ and the stars. To be conservative, we assumed all 100 stars to have a mass of 1\,\msun.  The code we used is described in \cite{2011MNRAS.416..618C}, the results of which have been verified by full SPH simulations \citep{2011ApJ...743...35C}.  We do not include a Galactic potential so that we can compare directly with our LMA solutions.  The encounter times are very similar, as are the encounter distances and velocities.  This suggests that N-body effects are unlikely to significantly affect the analysis of the origins of \oum\ (although it is possible that stars not in our sample, especially more massive ones, have some impact).

\subsection{Comparison to previous searches}

Several searches for a home system have been published using pre-\gdr{2} data. These searches, which we discuss here, predominantly used TGAS astrometry. This is less precise than \gdr{2} and limited searches to at most 0.3 million stars (those for which radial velocities were also available). These studies also assumed a purely gravitational trajectory for \oum.  The difference of about 0.4\degree\ between the purely gravitational solution JPL~13 and \candsaa\ (Figure~\ref{fig:asymptote_solutions}) corresponds to a difference of 0.35\,pc after a journey of 50\,pc (for linear motions).

\cite{Dybczynski2018} compiled data from Simbad, which was mostly TGAS and Hipparcos astrometry, and RAVE-DR4 \citep{2013AJ....146..134K} and RAVE-DR5 \citep{2017AJ....153...75K} radial velocities. They found \object{HIP 3757} to be the closest encounter, with  $\tenc=-0.118$\,Myr, $\denc=0.0044$\,pc, $\venc=185$\,\kms, 
but they sensibly disregarded this as a likely home for \oum\ due to the large encounter velocity. 
Although the authors do not specify the exact origin of the data for this star, this large velocity is probably a consequence of adopting the large radial velocity of $210.04 \pm 34.00$\,\kms\ listed in RAVE-DR4 (and RAVE-DR5).\footnote{If we use this value together with \gdr{2} astrometry in the LMA, we also get a very close encounter: $\tenc=-0.117$\,Myr, $\denc=0.073$\,pc, $\venc=208$\,\kms.}  
Yet there is good reason to doubt the validity of this measurement: its uncertainty is large, various quality indicators in RAVE suggest it is unreliable, and the velocity itself is surprisingly high. The smaller radial velocity from \gdr{2} of $26.88 \pm 0.35$\,\kms\ leads to a slower encounter (and a larger separation). Interestingly, RAVE-DR5 lists a second radial velocity for this star, based on a different spectrum, of $26.50 \pm 2.20$\,\kms, with good quality metrics. This is more likely to be closer to the true value. \cite{Dybczynski2018} were lucky that the configuration of the relative motion of \oum\ and 
\object{HIP 3757} is such that even with an erroneously large radial velocity one still finds a close encounter.

We note in passing that when using the larger RAVE-DR4 radial velocity and Hipparcos astrometry, \cite{2015A&A...575A..35B} found \object{HIP 3757} to approach to 2.0--4.4\,pc (90\% CI) of the Sun about 0.11\,Myr ago. Equipped now with \gdr{2}, this result is no longer valid.

\cite{Feng2018} used TGAS, Hipparcos, RAVE, and XHIP to search for past encounters with \oum. Their closest encounter was \object{HIP 21553} at 1.1\,pc, 35\,\kms,  0.28\,Myr ago. This is \object{\gdr{2} 272855565762295680} and is the tenth closest encounter in our Table~\ref{tab:periStats}, with consistent encounter parameters. We do not consider it as a home candidate simply because we have better (closer and/or slower) ones. One redeeming feature of this star, however, is that it is one of the most recent encounters, meaning that the backwards integration of its (and \oum's) path to encounter is even less sensitive to misrepresentation of the potential. 
This is neither a sufficient nor necessary conditions for being a home candidate, however.
A potentially more interesting object in the study of \cite{Feng2018} is \object{HIP 104539}, which has a low nominal encounter velocity of 10.2~(90\% CI 3.3--17.2)\,\kms. Unfortunately it has a huge uncertainty on its encounter distance: the 
90\% CI is 0.25--55\,pc. 
This object is in \gdr{2}, but without a radial velocity, so was not considered in our study. 
These authors also mention that they found \object{HIP 3757} as a fast, close encounter, but give no details, and dismissed it on the grounds of having a noisy spectrum, perhaps a reference to the RAVE-DR4 one.

The closest encounter found in a TGAS-based study from \cite{Portegies2018} 
was \object{HIP 17288} (note that the \gdr{1} source ID they give in their Table 1 is incorrect).
The encounter distance of 1.34\,pc (with a range of 1.33--1.36\,pc) and velocity of 14.9\,\kms\ make it a reasonable candidate.  Using presumably the same data, \cite{Feng2018} found a similar nominal distance, but surprisingly get a much larger 90\% CI of 0.07--7.81\,pc. This star is in our full sample, but with $\denclma=10.1$\,pc it was not an encounter candidate. The orbital integration with the surrogates on this star gave $\dencmed=4.16$ (90\% CI 2.92--5.46)\,pc. Both this large value and spread confirm that this is a poor candidate.  The other parameters, $\tencmed=-6.91$\,Myr and $\vencmed=15.3$\,\kms, are consistent with \cite{Portegies2018}.
The very small \denc\ range they quote (which is not defined) is surely either uncharacteristic or is underestimated. They in any case dismiss this and their more distant encounters as being too distant or too fast to be home candidates.

\cite{Zhang2018} did not find any convincing home candidates among his sample of 68 stars from TGAS within 10\,pc of the Sun.

Finally, \cite{Zuluaga2018} describe a method of assigning an origin probability for a particular star based on its encounter distance and velocity. 
Their most probable candidate, based on Hipparcos data,
is \object{HIP 103749}, for which they find $\denc=1.75$\,pc (with a large uncertainty of up to 5 pc) and $\venc=12.0$\,\kms. This is \object{\gdr{2} 1873204223289046272}, which we found to have $\denclma=16.4$\,pc and $\venclma=14.8$\,\kms. The discrepancy may be a result of the significantly different proper motion in \gdr{2} compared to both Hipparcos and Hipparcos-2. Their three next best encounters occur at larger separations and velocities. The first of these 
was excluded from our full sample on account of poor astrometry ($u=38.7$). The second
is not in \gdr{2} (the reason is not obvious, as it is isolated and has $V=11.3$\,mag). The third, \object{HIP 3821} = \object{\gdr{2} 425040000959067008}, we find for solution \candsaa\ to have $\denc$\,=\,2.62--2.67\,pc and $\venc$\,=\,24.75--25.17\,pc (90\% CIs).

None of the studies cited in this section mentions
\object{HD 292249}, \object{home-3}, or \object{home-4}.

\section{Conclusions}
\label{sec:conclusions}

We have integrated the trajectories of \oum\ and stars in the \gdr{2} catalog back in time over several million years through a smooth Galactic potential to identify mutual close encounters. Of the 28 (from \gaia\ only) + 13 (Simbad RVs) stars which approach within 2\,pc of \oum, four are of particular interest, with details as follows (for the non-gravitational asymptote solution \candsaa).
\begin{itemize}
\item The closest encounter is with the M2.5 dwarf \object{HIP 3757} with encounter parameters 
\tenc\,=\,$-0.99$ (90\% CI $-1.01$ to $-0.96$)\,Myr, 
\denc\,=\,0.60 (90\% CI 0.53--0.67)\,pc, 
\venc\,=\,24.7 (90\% CI 24.1--25.2)\,\kms. 
\item A more distant but slower encounter is with the G5 dwarf \object{HD 292249} with
\tenc\,=\,$-3.76$ (90\% CI $-4.37$ to $-3.30$)\,Myr, 
\denc\,=\,1.60 (90\% CI 1.38--1.87)\,pc, 
\venc\,=\,10.7 (90\% CI 9.3--12.2)\,\kms.
\item Intermediate between these is the star (probably a K dwarf) \object{home-3} with 
\tenc\,=\,$-6.30$ (90\% CI $-6.99$ to $-5.68$)\,Myr, 
\denc\,=\,1.01 (90\% CI 0.30--1.95)\,pc, 
\venc\,=\,14.4 (90\% CI 13.1--15.9)\,\kms,
\item and the M5 dwarf \object{home-4} with
\tenc\,=\,$-1.14$ (90\% CI $-1.57$ to $-0.91$)\,Myr, 
\denc\,=\,0.88 (90\% CI 0.70--1.18)\,pc, 
\venc\,=\,17.9 (90\% CI 13.0--22.5)\,\kms.
\end{itemize}
The confidence intervals reflect the uncertainties in the 6D (space and velocity) coordinates of both the stars (which dominate) and \oum. The different non-gravitational solutions for the latter give rise to changes in the encounter parameters on the order of the uncertainties reported for distance, but much less for velocity. We find somewhat different encounter parameters for specific stars compared to earlier studies, partly due to our use of a more accurate, non-gravitational solution for \oum's trajectory, but mostly because of our use of more precise data from \gdr{2} compared to that in TGAS. Use of \gdr{2} has also increased the number of stars we could consider by a factor about 20.

The discovery of a close encounter by \oum\ to a specific star could mean that \oum\ was ejected from that star's planetary system. A plausible ejection mechanism is scattering by giant planets in that system, although to achieve an ejection velocity of 25\,\kms\ or even 10\,\kms\ would require a very close (and thus rare) planet--\oum\ encounter. None of our home star candidates is known to harbor giant planets.
Higher ejection velocities are more readily attained in a binary star system.
Another potential mechanism is ejection by close passage of a second star. Although we have found one star which has a small probability (20\%) of passing close ($<0.5$\,pc) to \object{home-4} at the right time, it is still very unlikely to have been close enough to have caused the ejection.

It is in fact unsurprising that we find some stars encountering within 2\,pc of \oum\ over the past few million years. 
The recent study by \cite{2018A&A...616A..37B} inferred, after correcting for selection effects, that of order 20 stars (of all masses) pass within 1\,pc of the Sun every Myr (most of which we would not identify in \gdr{2}).
Thus merely having a close encounter with \oum\ is not sufficient to be a home candidate star: given the plausible ejection mechanisms, a good candidate is one which also has a low relative velocity relative to \oum.

The main limitation of our study is the limited number of stars. We require 6D phase space information, and \gdr{2} provides this for ``only'' 7 million (although before \gaia\ we had only 0.1\, million). This is a magnitude limited sample ($\gmag\lesssim 14$) which therefore excludes many low mass and more distant stars (plus very bright stars). If we assumed that \oum\ could originate from any of the more than 35 million stars currently lying within 500\,pc of the Sun, we get a sense of the prior probability that we would have found the home star. The situation should improve significantly with \gdr{3}, when radial velocities should be published for an order of magnitude more stars.

Uncertainties in integrated orbits will of course grow as we go further back in time, and this too will ultimately limit our ability to identify home stars.  We adopted a smooth, time-independent potential, without the bar or spiral arms. While none of these are likely to be a significant limitation on the short time scales (few Myr) and distance scales ($\lesssim 100$\,pc) considered here, they will ultimately become a limiting factor, even with arbitrarily precise astrometry. 

The search for \oum's home continues.

\acknowledgments

This study has used data from the European Space Agency (ESA) mission \gaia\ (\url{http://www.cosmos.esa.int/gaia}), processed by the \gaia\ Data Processing and Analysis Consortium (DPAC, \url{http://www.cosmos.esa.int/web/gaia/dpac}). Funding for the DPAC has been provided by national institutions, in particular the institutions participating in the \gaia\ Multilateral Agreement. Funding for RAVE (www.rave-survey.org) has been provided by institutions of the RAVE participants and by their national funding agencies. We are grateful for the use of the Simbad object database and VizeR catalog service, both developed and provided by CDS, Strasbourg. DF conducted this research at the Jet Propulsion Laboratory, California Institute of Technology, under a contract with NASA. KJM acknowledges support through awards from the National Science Foundation AST1617015 and through HST programs GO/DD-15405 and -15447 provided by NASA through a grant from the Space Telescope Science Institute, which is operated by the Association of Universities for Research in Astronomy, Inc., under NASA contract NAS 5-26555. 

We thank Eric Mamajek (JPL) for helpful discussions during this analysis.


\bibliographystyle{aasjournal}
\bibliography{stellar_encounters,gaia,oumuamua,other}

\end{document}

%% file: figures/cands41plushome4_encounters.tex
$^{\dag}$2525688198820543360 &     -986 &    -1010 &     -964 &    0.599 &    0.527 &    0.669 &    24.7 &    24.1 &    25.2 &    40.23 &     0.06 &   218.22 &     0.20 &    26.9 &     0.3 & 11.14 &  9.16 &  2.17 \myeol 
 3174397001191766400 &    -3782 &    -5801 &    -2750 &    0.728 &    0.215 &    3.442 &    40.9 &    26.8 &    55.9 &     6.36 &     0.02 &    19.93 &     0.03 &    62.3 &     9.0 & 13.23 &  7.25 &  1.98 \myeol 
 4399766976222478336 &    -1777 &    -1805 &    -1748 &    0.775 &    0.589 &    0.983 &    47.0 &    46.3 &    47.7 &    11.72 &     0.04 &    51.91 &     0.07 &    31.3 &     0.4 & 11.13 &  6.48 &  1.17 \myeol 
 4531404425323318656 &    -2456 &    -2591 &    -2326 &    0.805 &    0.217 &    1.869 &    48.9 &    48.5 &    49.2 &     8.16 &     0.26 &    10.05 &     0.52 &    23.0 &     0.2 & 10.24 &  4.80 &  1.39 \myeol 
 6029459365812535296 &    -1285 &    -1321 &    -1251 &    0.853 &    0.478 &    1.211 &    95.3 &    93.0 &    97.6 &     7.99 &     0.06 &    40.06 &     0.11 &    85.3 &     1.4 & 10.61 &  5.12 &  0.82 \myeol 
 2764431961787219328 &    -4029 &    -4463 &    -3650 &    0.989 &    0.561 &    1.750 &    39.7 &    35.9 &    43.7 &     6.19 &     0.04 &    31.23 &     0.08 &    28.9 &     2.5 & 12.84 &  6.80 &  1.33 \myeol 
$^\P$2502921019565490176 &    -6300 &    -6989 &    -5677 &    1.006 &    0.300 &    1.954 &    14.4 &    13.1 &    15.9 &    11.13 &     0.04 &    57.73 &     0.13 &    23.3 &     0.9 & 11.99 &  7.23 &  1.55 \myeol 
 1774513537034328704 &     -231 &     -238 &     -224 &    1.026 &    0.962 &    1.090 &   419.4 &   412.3 &   426.1 &    10.12 &     0.16 &    35.72 &     0.39 &   402.2 &     4.1 & 15.40 & 10.42 &  2.64 \myeol 
 1815070015563299200 &    -2000 &    -3096 &    -1464 &    1.058 &    0.746 &    1.577 &    46.8 &    30.3 &    64.2 &    10.43 &     0.04 &    26.87 &     0.08 &    23.5 &    10.3 & 11.09 &  6.18 &  1.20 \myeol 
  272855565762295680 &     -282 &     -284 &     -280 &    1.086 &    1.078 &    1.095 &    34.2 &    34.0 &    34.5 &   100.81 &     0.06 &   563.96 &     0.25 &    33.7 &     0.2 &  7.95 &  7.97 &  1.81 \myeol 
 4691195659898343936 &    -1219 &    -1261 &    -1179 &    1.138 &    1.092 &    1.186 &    89.9 &    87.0 &    93.0 &     8.93 &     0.01 &    40.14 &     0.04 &   105.1 &     1.8 & 13.14 &  7.90 &  1.72 \myeol 
 3315373564607612416 &     -637 &     -658 &     -617 &    1.228 &    1.177 &    1.282 &    69.7 &    67.4 &    72.0 &    22.01 &     0.04 &    81.00 &     0.08 &    90.8 &     1.4 & 11.60 &  8.31 &  1.93 \myeol 
 6644271495897847808 &    -2300 &    -2329 &    -2271 &    1.334 &    0.843 &    1.966 &    61.6 &    61.2 &    62.0 &     6.91 &     0.05 &    38.77 &     0.07 &    60.7 &     0.2 &  9.47 &  3.66 &  0.66 \myeol 
 6915359008301010944 &    -2762 &    -2937 &    -2615 &    1.337 &    1.158 &    1.548 &    32.3 &    30.4 &    34.1 &    11.00 &     0.04 &    46.24 &     0.09 &    15.3 &     1.1 & 12.91 &  8.12 &  1.85 \myeol 
 4565244438273756928 &    -3441 &    -3619 &    -3287 &    1.360 &    0.649 &    2.352 &    33.3 &    31.9 &    34.8 &     8.57 &     0.13 &    20.79 &     0.25 &     9.0 &     0.8 &  7.40 &  2.07 &  0.73 \myeol 
  951511177924715392 &    -3463 &    -3513 &    -3415 &    1.367 &    0.624 &    2.138 &    33.3 &    32.9 &    33.7 &     8.47 &     0.04 &    45.29 &     0.09 &    40.0 &     0.2 & 10.73 &  5.37 &  0.94 \myeol 
  211810233512673792 &     -489 &     -508 &     -470 &    1.410 &    1.356 &    1.469 &    26.5 &    25.5 &    27.6 &    75.04 &     0.04 &   437.96 &     0.11 &    30.2 &     0.6 &  9.31 &  8.69 &  2.14 \myeol 
 1252717513566583552 &    -4489 &    -4681 &    -4316 &    1.419 &    0.735 &    2.372 &    34.5 &    33.2 &    35.7 &     6.45 &     0.04 &    29.79 &     0.08 &    17.7 &     0.8 & 11.26 &  5.31 &  1.01 \myeol 
 4319754583064325248 &    -3738 &    -3937 &    -3551 &    1.490 &    1.281 &    1.712 &    25.6 &    24.3 &    27.0 &    10.20 &     0.03 &    20.73 &     0.07 &     1.1 &     0.8 & 10.42 &  5.46 &  0.98 \myeol 
$^{\ddag}$3107000885484340224 &    -3765 &    -4361 &    -3300 &    1.605 &    1.377 &    1.868 &    10.7 &     9.3 &    12.2 &    24.15 &     0.05 &    71.11 &     0.10 &    33.1 &     0.9 &  9.09 &  6.01 &  1.09 \myeol 
 4955574948784195840 &    -2146 &    -2209 &    -2082 &    1.676 &    1.461 &    1.901 &    75.5 &    73.4 &    77.7 &     6.07 &     0.02 &    29.07 &     0.03 &    89.1 &     1.4 & 12.98 &  6.89 &  1.41 \myeol 
 2264912868532666240 &    -7692 &    -8032 &    -7365 &    1.697 &    0.988 &    2.462 &    14.6 &    14.0 &    15.2 &     8.91 &     0.03 &    31.93 &     0.08 &    -6.5 &     0.4 &  9.80 &  4.55 &  0.80 \myeol 
 1820385295296465280 &    -8776 &    -9737 &    -7980 &    1.724 &    0.642 &    3.211 &    30.7 &    27.8 &    33.7 &     3.64 &     0.04 &     8.05 &     0.07 &     6.3 &     1.7 & 12.61 &  5.41 &  0.97 \myeol 
 2425971152835667072 &    -5868 &    -6038 &    -5701 &    1.736 &    1.361 &    2.157 &    17.2 &    16.7 &    17.6 &    10.12 &     0.04 &    55.55 &     0.11 &    17.8 &     0.3 & 10.94 &  5.97 &  1.04 \myeol 
 2984656203733095936 &    -1296 &    -1325 &    -1267 &    1.773 &    1.706 &    1.844 &    28.3 &    27.7 &    29.0 &    26.63 &     0.04 &    69.13 &     0.10 &    52.3 &     0.4 & 11.55 &  8.68 &  2.11 \myeol 
 6142461810682406656 &    -1598 &    -1661 &    -1532 &    1.930 &    1.734 &    2.139 &    41.9 &    40.2 &    43.6 &    14.63 &     0.06 &    81.33 &     0.11 &    48.5 &     1.1 & 12.47 &  8.29 &  2.16 \myeol 
 3386274609851856896 &    -1540 &    -1567 &    -1515 &    1.960 &    1.790 &    2.166 &    93.2 &    92.2 &    94.2 &     6.81 &     0.05 &    33.80 &     0.12 &   105.5 &     0.6 &  9.40 &  3.56 &  0.77 \myeol 
 3234412606443085824 &    -1183 &    -1212 &    -1156 &    1.991 &    1.911 &    2.078 &    16.3 &    15.9 &    16.7 &    50.42 &     0.11 &   191.27 &     0.19 &    36.3 &     0.2 &  7.54 &  6.05 &  1.21 \myeol 
\hline
$^{\sharp}$ 3666992950762141312 &    -1141 &    -1567 &     -905 &    0.875 &    0.704 &    1.177 &    17.9 &    13.0 &    22.5 &    48.03 &     0.10 &   245.97 &     0.19 &     7.5 &     2.9 & 13.26 & 11.67 &  3.27 \myeol 